\documentclass[reprint,superscriptaddress,amsmath,amssymb,aps,prb]{revtex4-1}
\usepackage{graphicx}
\usepackage{placeins}
\usepackage{bm}

\begin{document}

\title{Circular polarization reversal of half-vortex cores in polariton condensates}

\author{Matthias Pukrop}
\affiliation{%
 Department of Physics and CeOPP, Universit\"at Paderborn, Warburger Stra{\ss}e 100, 33098 Paderborn, Germany
}
\author{Stefan Schumacher}
\affiliation{%
 Department of Physics and CeOPP, Universit\"at Paderborn, Warburger Stra{\ss}e 100, 33098 Paderborn, Germany
}%
\affiliation{College of Optical Sciences, University of Arizona, Tucson, AZ 85721, USA}
\author{Xuekai Ma}
\affiliation{%
 Department of Physics and CeOPP, Universit\"at Paderborn, Warburger Stra{\ss}e 100, 33098 Paderborn, Germany
}

\date{\today}

\begin{abstract}
Vortices are topological objects carrying quantized orbital angular momentum and have been widely studied in many physical systems for their applicability in information storage and processing. In systems with spin degree of freedom the elementary excitations are so called half-vortices, carrying a quantum rotation only in one of the two spin components. We study the spontaneous formation and stability of localized such half-vortices in semiconductor microcavity polariton condensates, non-resonantly excited by a linearly polarized ring-shaped pump. The TE-TM splitting of optical modes in the microcavity system leads to an effective spin-orbit coupling, resulting in solutions with discrete rotational symmetry. The cross-interaction between different spin components provides an efficient method to realize all-optical half-vortex core switching inverting its circular polarization state. This switching can be directly measured in the polarization resolved intensity in the vortex core region and it can also be applied to higher order half-vortex states.
\end{abstract}

\maketitle

\begin{section}{Introduction}
Vortices carry quantized orbital angular momentum (OAM), also known as topological charge, and have been widely studied in many physical systems. Considering spin degree of freedom the elementary excitations are so called half-vortices (HV). A HV refers to a vortex state carrying quantized OAM in one component of a spinor system, while in the other component there is a fundamental mode (FM) without OAM, distinguishing them from full-vortices (FV) with the same OAM in both components and spin-vortices (SV) with the oppotsite OAM in both components. HVs have been predicted and observed in superconductors~\cite{luk1995magnetic,jang2011observation}, $^3$He superfluids~\cite{PhysRevLett.55.1184,PhysRevLett.117.255301}, antiferromagnetic atomic condensates~\cite{PhysRevA.85.023606,PhysRevLett.112.180403,PhysRevLett.115.015301,seo2016collisional}, as well as exciton-polariton condensates~\cite{rubo2007half,lagoudakis2009observation,manni2012dissociation}. 

Exciton-polaritons are half-light half-matter quasi-particles formed in semiconductor microcavities with strong coupling of quantum-well (QW) excitons and cavity photons. The hybrid nature enables polaritons to be optically excited through their photonic part, and to strongly interact with each other through their excitonic part, with the latter giving rise to strong optical nonlinearities. With the polarization dependence of microcavity polariton excitations, a complex interplay of TE-TM cavity-mode splitting in the linear optical regime and spin-dependent exciton-exciton interactions in the nonlinear regime arises. The latter leads to repulsive polariton-polariton interaction for two polaritons in the same circular polarization component and, for the spectral range considered here, an attractive (cross-species) interaction between polaritons with opposite circular polarization~\cite{PhysRevB.58.7926,PhysRevB.82.075301}. A number of interesting phenomena regarding spinor polaritons in semiconductor microcavities have been observed, such as the optical spin-Hall effect~\cite{kavokin2005optical,leyder2007observation,Lafont_Controlling_the_optical_spin_hall_effect_APL2017},  formation of Skyrmions~\cite{flayac2013transmutation}, and HVs~\cite{rubo2007half,lagoudakis2009observation,manni2012dissociation}.

HVs in polariton condensates can be excited by resonant pumping through direct imprinting of OAM~\cite{dominici2015vortex} or form spontaneously through disorder scattering~\cite{PhysRevLett.101.187401}. Under non-resonant excitations they can also be formed spontaneously due to sample defects ~\cite{lagoudakis2009observation,PhysRevB.91.085413}. The subsequent optical HVs emitted from the microcavity are defined by a circular polarized light field in the core region with frequencies, in general, lower than those of FVs. However, until now localized HVs, spontaneously formed under non-resonant excitation via linearly polarized pumps have not been investigated, although localized vortices have been intensively investigated in scalar polariton condensates (considering only one circular polarization component)~\cite{dall2014creation,ma2016incoherent,ma2018vortex,kwon2019direct,ma2019realization}. Localized vortices are beneficial as they are spatially trapped and can also carry information with higher topological charges. Using a ring-shaped trap a different type of half-quantum circulation can be realized characterized by a spin flip and linear polarization reversal on opposite sides of the ring~\cite{liu2015new}. Recent work shows that due to a spatial dislocation of the pump ring centers in different polarization components a state with a fundamental mode in one component and a topological state oscillating between $m{=}{\pm} 1$ in the other component can be generated~\cite{yulin2016spontaneous}.

\begin{figure}[t]
		\centering
		\includegraphics[width=0.49 \textwidth]{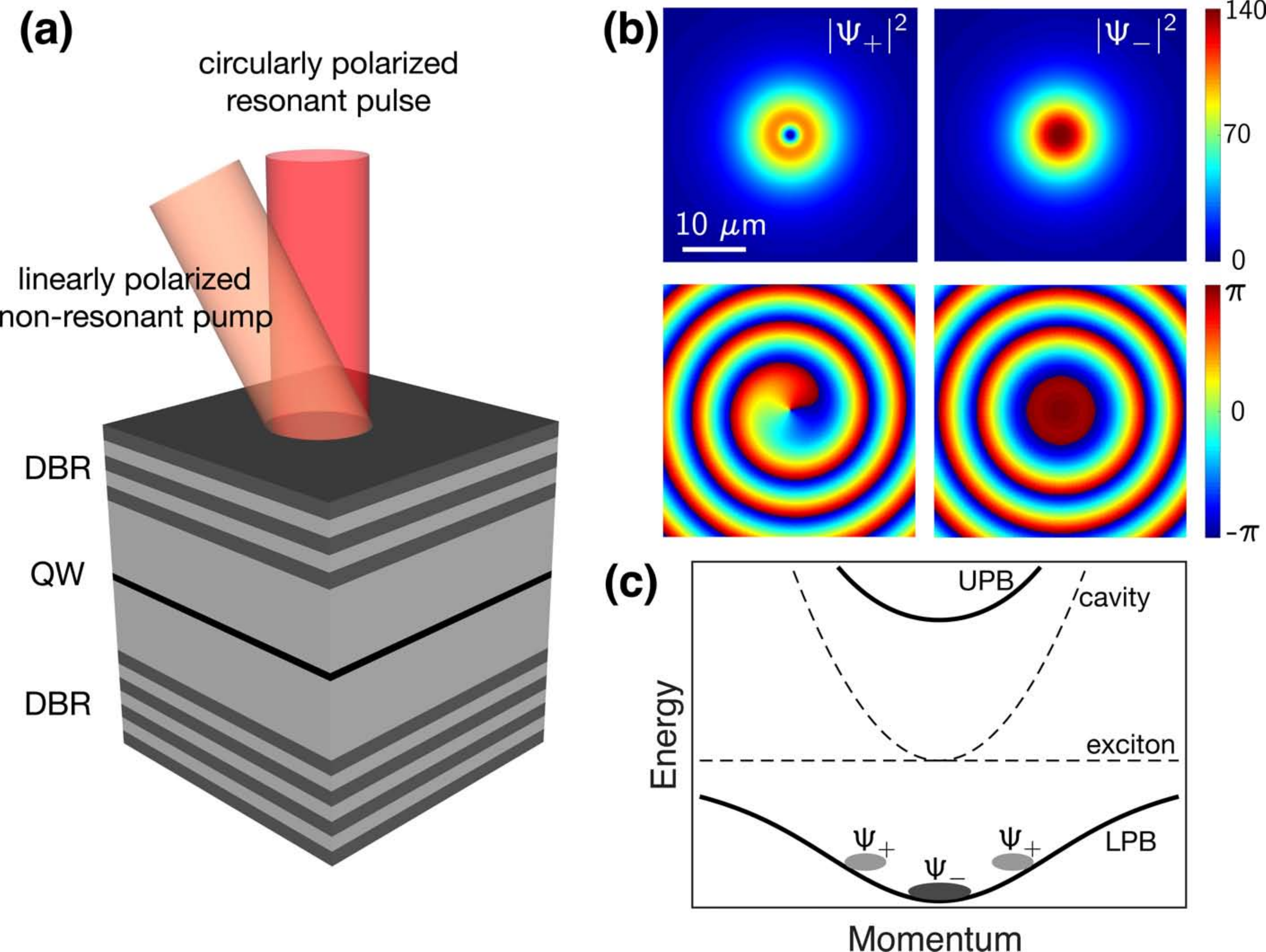}
		\caption{(a) Sketch of the planar semiconductor microcavity excited by a linearly polarized non-resonant ring pump. A circularly polarized resonant ring pulse is used to achieve vortex core switching. (b) Density ($\mu m^{{-}2}$) and phase of the two circular polarization components for a half-vortex $(-1,0)$ with $\Delta_{\mathrm{LT}}{=}0$. Pump radius is $\mathrm{w}_p{=}4.5~\mu\mathrm{m}$. (c) Sketch of the normal-mode splitting of the polariton dispersion into upper (UPB) and lower (LPB) polariton branch (\textit{solid}), as well as the bare cavity and exciton dispersion (\textit{dashed}). Also schematically indicated are the condensate components for the half-vortex state.}
		\label{fig:microcavity}
\end{figure}

In this work, we report localized HVs in polariton condensates, excited by a ring-shaped non-resonant pump. Importantly, we find an all-optical method to switch the OAM of the HV from one component to the other due to the attractive cross-interaction by applying a resonant circularly polarized pulse. Such kind of switching reverses the circular polarization direction of the emitted light from the vortex core, realizing optical vortex core switching in analogy to the switching dynamics widely studied in magnetic systems, promising applications in data-storage devices~\cite{schneider2001magnetic,xiao2006dynamics,hertel2007ultrafast,yamada2007electrical}. In contrast to our previous work where the topological charge of a localized vortex is inverted by a non-resonant pulse~\cite{ma2019realization}, here we switch between left and right circularly polarized HVs using a resonant pulse. This polarization reversal happens due to the resonant pulse which directly imprints a vortex charge in its corresponding spin component, while destabilizes the prexisting vortex in the opposite spin component. It is worth mentioning that in scalar condensates the phase of a vortex needs to be measured to confirm the sign of its topological charge (or rotation direction) as a binary information because the density in the core region is zero. The HV states in spinor condensates lead to a nonzero density in the vortex core with pure circular polarization, which provides an easy way to probe the binary polarization information by only measuring the polarization resolved density in the core region. Higher-order HV states with the vortex component carrying a higher topological charge $|m|{>}1$ are also studied and we show that they can be switched either to a lower energy FV or SV state or to an oppositely polarized HV state, enabling multi-level switching configurations. Besides HVs the linearly polarized ring-shaped excitation setup can also support FVs as well as both FMs. Considering the TE-TM splitting, there is a debate about the stability of HVs in the conservative system~\cite{flayac2010topological,solano2010comment,flayac2010reply}. Here, we find that HVs are still stable in the presence of loss and gain, even if the TE-TM splitting value is significant. In agreement with previous work, the gain acts stabilizing for these vortices~\cite{lobanov2011stable}. With TE-TM splitting, the profiles of the HVs become warped instead of being centrosymmetric~\cite{toledo2014warping}. The paper is organized as follows. In Section \ref{sec:model} we introduce the model and discuss the multistability of different localized vortex solutions in a spinor polariton field excited by a ring-shaped non-resonant pump. In Section \ref{sec:switching} we present a method to switch between different polarized HVs, discuss the physical mechanism, and show a representative example. In Section \ref{sec:higher} we extend the switching scheme to higher order states and additionally discuss the effect of TE-TM splitting on the solutions' symmetries.
\end{section}

\begin{section}{Model}\label{sec:model}
We use the driven-dissipative Gross-Pitaevskii equation coupled to the density of an incoherent reservoir to describe the dynamics of the polariton condensate close to the bottom of the lower polariton branch in the circular polarization basis~\cite{PhysRevLett.109.036404}
\begin{align}
\mathrm{i}\hbar\dot{\Psi}_{\pm}=& \bigg[H-\mathrm{i}\hbar\frac{\gamma_c}{2}+g_c|\Psi_{\pm}|^2-g_{\mathrm{x}}|\Psi_{\mp}|^2 \label{eq:GPE_psi} \\
& +(g_r+\mathrm{i}\hbar\tfrac{R}{2})n_{\pm} \bigg]\Psi_{\pm}+H^{\pm}\Psi_{\mp}+E_{\pm} \nonumber \\
\dot{n}_{\pm}=&\left(-\gamma_r-R|\Psi_{\pm}|^2 \right)n_{\pm}+P_{\pm} \label{eq:GPE_n}\,.
\end{align}
Here the indices $\pm$ mark the right/left circular polarization components. $H{=}{-}\tfrac{\hbar^2}{2m_{\mathrm{eff}}}\nabla^2$ is the free particle Hamiltonian with effective polariton mass $m_{\mathrm{eff}}$. $\gamma_c$ and $\gamma_r$ are the decay rates of the condensate and the reservoir with the relation $\gamma_r=1.5\gamma_c$. We would like to note that the ratio we used here~\cite{dall2014creation,Roumpos2011} indicates that the reservoir decays much faster than that observed in experiments. The reason is that the simple GP model does not include the energy relaxation~\cite{PhysRevB.82.245315} happening during the condensation. It has been shown that the energy relaxation significantly influences the dynamics of the condensate when the pump intensity is much larger than the threshold, but its influence becomes weaker as the pump intensity decreases, especially when the pump power is just above the threshold~\cite{PhysRevB.93.121303}. Interaction strength between polaritons with parallel (opposite) spins is described by $g_c$ ($g_\mathrm{x}$). Condensation rate and condensate-reservoir interaction are given by $R$ and $g_r$. $H^{\pm}{=}\tfrac{\Delta_{\mathrm{LT}}}{k_{\mathrm{LT}}^2}\left(\mathrm{i}\partial_x \pm \partial_y \right)^2$ describes the effect of TE-TM splitting intrinsically present in microcavities~\cite{PhysRevB.59.5082}. Parameters used are explicitly given below in Ref.~\cite{parameters}. $E_{\pm}$ represents an optional coherent control beam. The system is excited non-resonantly with an $x$-linearly polarized continuous-wave ring pump with radius $\mathrm{w}_p$ and intensity $P_0=5\cdot P_{\mathrm{thr}}$ with condensation threshold $P_{\mathrm{thr}}$, i.e.
\begin{equation}
P_+=P_-=P_0~\tfrac{\mathbf{r}^2}{\mathrm{w}_p^2}~\mathrm{e}^{-\mathbf{r}^2/\mathrm{w}_p^2}.
\end{equation}

Vortex states can be characterized by two topological charges, denoted as $(m_+$,$m_-)\in\mathbb{Z}^2$ hereinafter, describing the rotation $2\pi m_{\pm}$ of the phases of the corresponding components $\Psi_{\pm}$ while encircling the vortex core center. 
\begin{figure}[t]
		\centering
		\includegraphics[width=0.48 \textwidth]{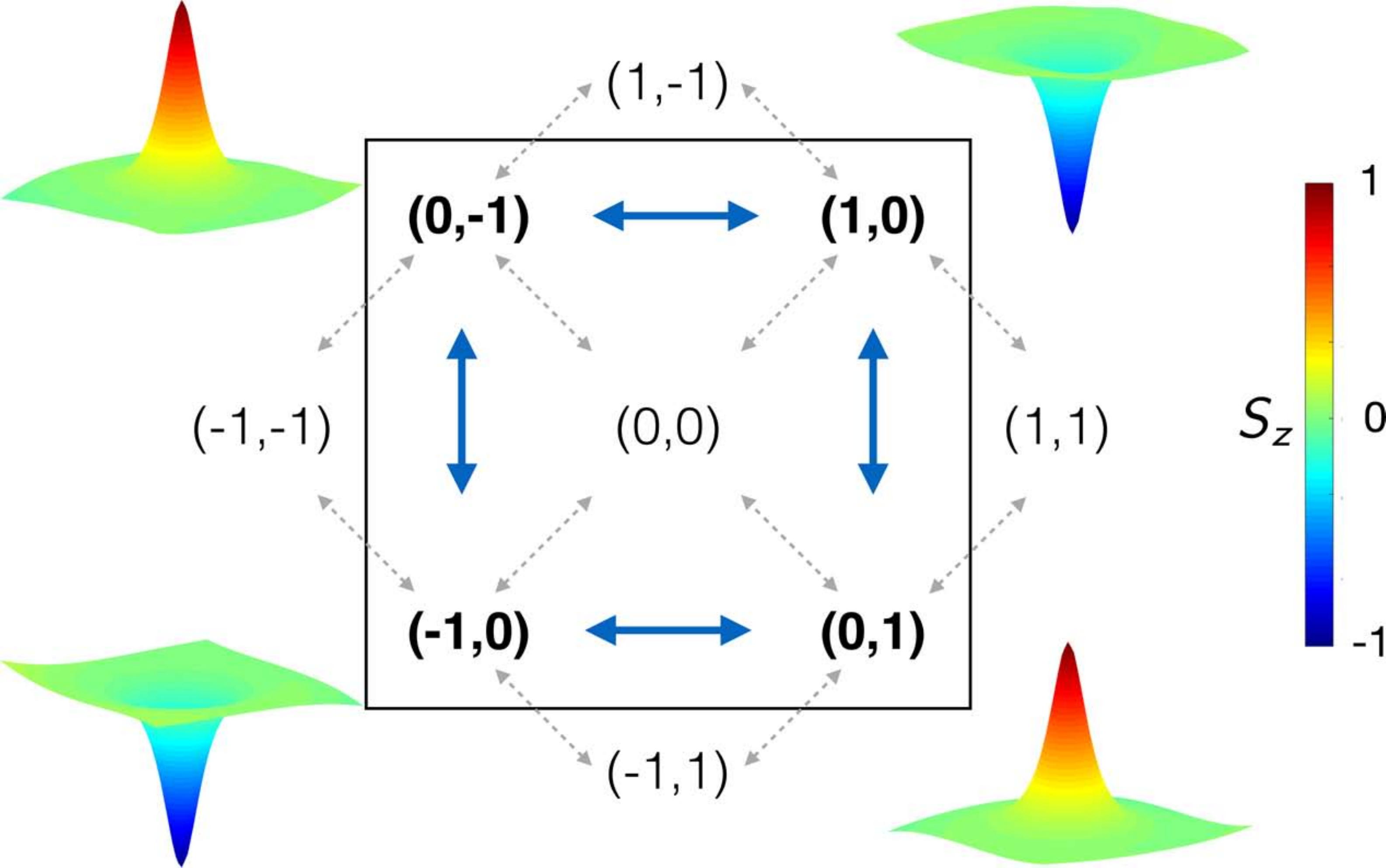}
		\caption{Schematic arrangement of the four elementary half-vortices (\textit{bold}) together with the next nearest full-vortices and fundamental-mode states. Numbers in brackets refer to the topological charges $(m_+,m_-)$ of the two polarization components. \textit{Blue} arrows indicate possible vortex core switching. \textit{Grey} arrows represent OAM imprinting in one component without changing the other one. Corners show pseudospin component $S_z$ (degree of circular polarization) in real space near the core region in the cavity plane for the four elementary HVs.}
		\label{fig:HV_switching}
\end{figure}
In the present driven-dissipative system described by Eqs.~\eqref{eq:GPE_psi} and \eqref{eq:GPE_n} excited by an incoherent ring-shaped pump, localized vortex solutions are stabilized due to the balance between outgoing propagation of the condensate and the confinement of the reservoir-induced potential \cite{dall2014creation,lobanov2011stable}. Generally, the vortices can be stable only when their radii are smaller than the reservoir potential. This balance is achieved by reshaping the reservoir-induced potential through condensate-reservoir interactions and induced scattering rate. The reshaping of the reservoir potential is more pronounced if the vortex radius increases by carrying higher topological charge. In this scenario, vortex states may be unstable when their radius exceeds the radius of the potential. In our numerical simulations based on Eqs. \eqref{eq:GPE_psi} and \eqref{eq:GPE_n} in 2D real space we find stable HV solutions which can either arise from random white noise or be initialized by corresponding initial conditions (coherent imprinting of a HV). They remain stable if exposed to random noise (with 5$\%$ of the solution magnitude) during the time evolution, evidencing that the solutions are very robust. HVs are states where a $m{=}{\pm} 1$ charged vortex exists in one component while there is no vortex ($m{=}0$) in the other component. Figure \ref{fig:microcavity}(b) shows densities and phases of the HV $(-1,0)$ for the two polarization components. Also schematically shown are the spectra of the two condensate components for the HV state on the lower polariton branch in Fig.~\ref{fig:microcavity}(c), indicating that the non-vortex and vortex state have different energies, representing ground and first excited state of the system. In contrast to FVs and SVs, HVs have a singularity in only one polarization component. In the center of the HV core, this results in pure right circular polarization for \textit{right} HVs $(0,\pm 1)$ and pure left circular polarization for \textit{left} HVs $(\pm 1,0)$. The corners of Fig. \ref{fig:HV_switching} show the pseudospin component $S_z{=}(|\Psi_+|^2{-}|\Psi_-|^2)/(|\Psi_+|^2{+}|\Psi_-|^2)$ for the four elementary HVs in the core region which is directly related to the degree of circular polarization of the emitted light \cite{dzhioev1997determination}. Besides the four HVs (\textit{bold} in Fig. \ref{fig:HV_switching}), we find that FVs with $|m_{\pm}|{=}1$ and FM with $m_{\pm}{=}0$ are also stable for the same parameters and under the same excitation conditions. Including TE-TM splitting does not affect the stability of the solutions but in general lifts the degeneracy of the FVs~\cite{PhysRevLett.115.246401,Zambon2019} as well as breaks the cylindrical symmetry~\cite{solano2010comment} and only a reduced discrete symmetry is conserved depending on the difference of the two topological charges $\Delta m{=}m_+{-}m_-$, resulting in the reshaping of the density distributions (see Section \ref{sec:higher} below). The orientations of the patterns rotate over time with periods of about $400~\mathrm{ps}$ for the parameters in the present work. The period depends on the TE-TM splitting strength. We note that an alternative description of HVs in the linear polarization basis uses the polarization and phase rotation winding numbers $(k,m)$~\cite{rubo2007half,Manni2013}, with the following relation to our notation: $k{=}(m_-{-}m_+)/2$, $m{=}(m_+{+}m_-)/2$.
\end{section}

\begin{section}{Half-vortex Switching}\label{sec:switching}
We demonstrate reversible on-demand switching between left and right HVs using a coherent control pulse in only one circular polarization component
\begin{equation}
E_{\pm}=E_0~\mathbf{r}^2~\mathrm{e}^{-\mathbf{r}^2/\mathrm{w}_c^2}~\mathrm{e}^{\mathrm{i}m_c\varphi}~\mathrm{e}^{-\mathrm{i}\omega_c t}\mathrm{e}^{-(t-t_0)^2/\mathrm{w}_t^2}.
\end{equation}
The control pulse is also ring-shaped with radius $\mathrm{w}_c$ and carries a topological charge $m_c$. The frequency $\omega_c$ is chosen to be nearly resonant with the component where the control pulse is applied.

\begin{figure}[t]
		\centering
		\includegraphics[width=0.49 \textwidth]{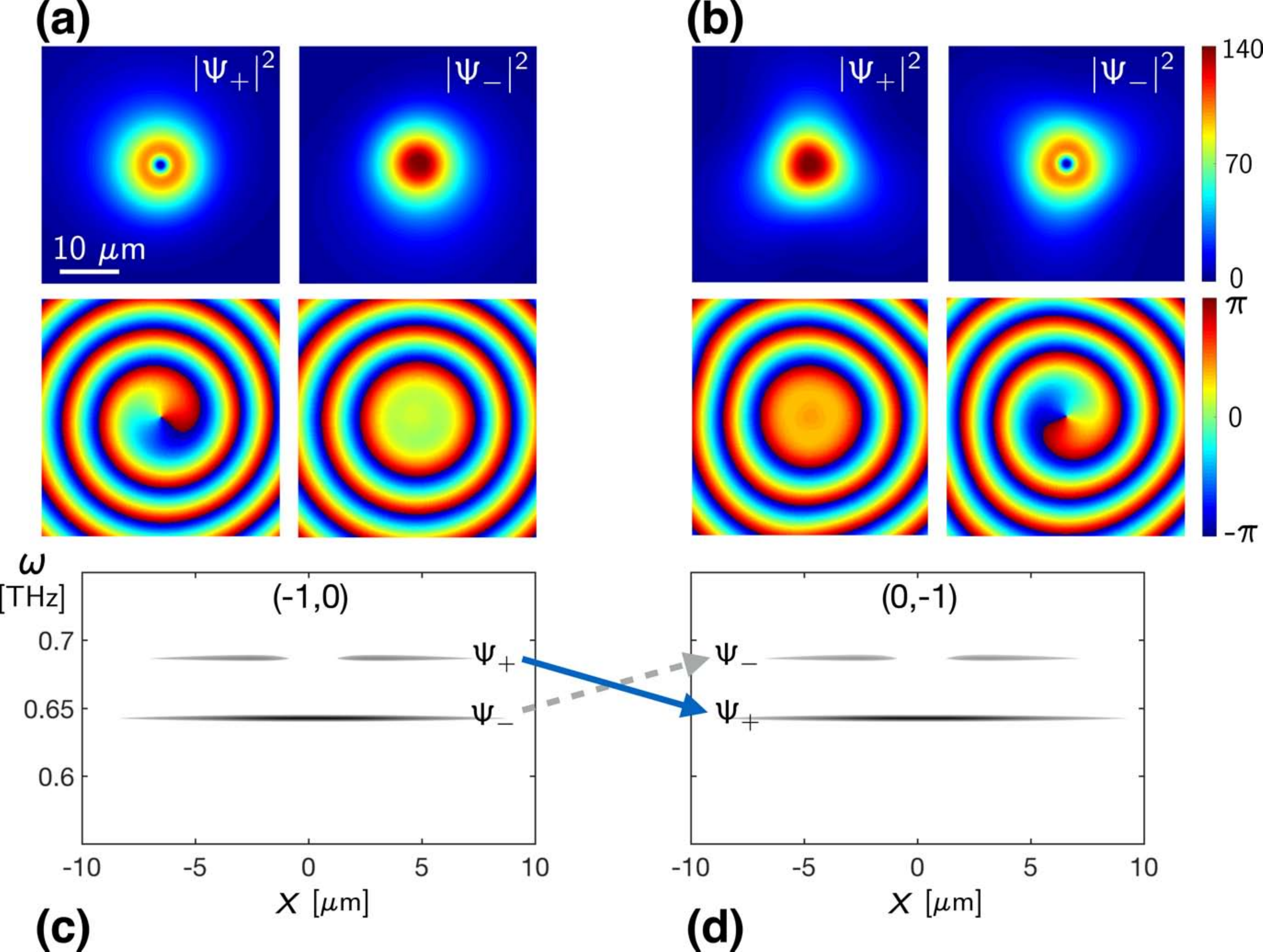}
		\caption{Vortex core switching from $(-1,0)$ to $(0,-1)$. Densities ($\mu m^{{-}2}$) and phase profiles of (a) the initial and (b) the final state. (c,d) Spectra in real space, corresponding to (a,b), respectively. \textit{Grey dashed} arrow represents imprinting of a vortex state in the $\Psi_{-}$ component due to the resonant control beam, while the \textit{blue} arrow shows the simultaneous transition into the ground state of the $\Psi_{+}$ component. Pump radius is $\mathrm{w}_p{=}4.5~\mu\mathrm{m}$. Control pulse parameters are: $E_0=2.5{\times} 10^{-3}~\mu\mathrm{m}^{-3}$, $\mathrm{w}_t{=}35~\mathrm{ps}$, $\mathrm{w}_c{=}6~\mu\mathrm{m}$.}
		\label{fig:HV_switching_example}
\end{figure}

\begin{figure}[t]
		\centering
		\includegraphics[width=0.43 \textwidth]{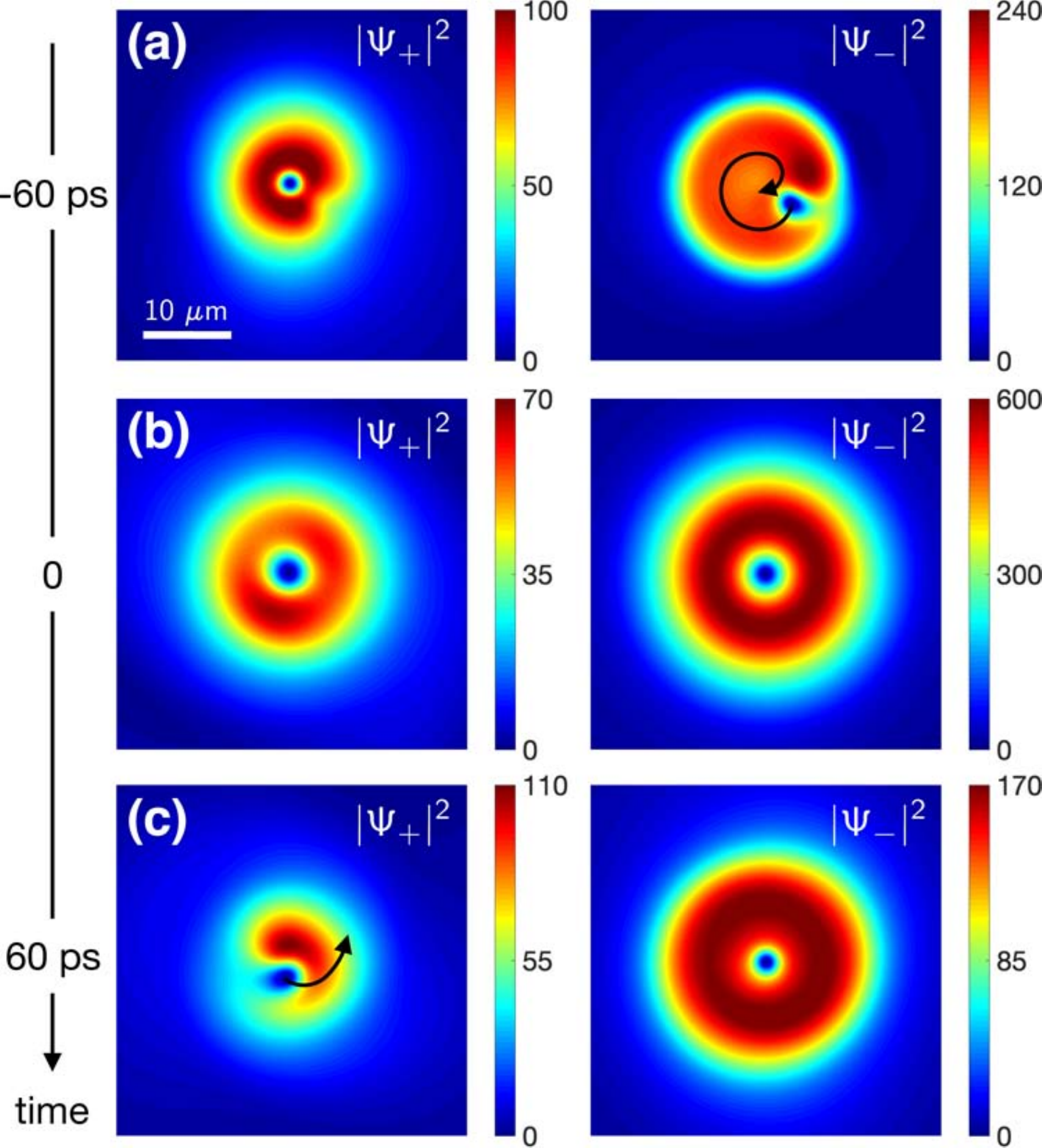}
		\caption{Densities of both components during the switching process at three different times [(a)-(c)] for the example shown in Fig. \ref{fig:HV_switching_example}. \textit{Black arrows} represent vortex trajectories. The resonant pulse applied to $\Psi_-$ has an Gaussian envelope in time centered at $t{=}0$ with width $\mathrm{w}_t{=}35~\mathrm{ps}$. (a) Vortex in $\Psi_-$ spirals in due to the resonant pulse. (b) Pulse maximum is reached. Vortex in $\Psi_+$ starts to precess demonstrating it has become unstable. (c) Vortex in $\Psi_+$ spirals out of the condensate (\textit{black arrow}).}
		\label{fig:switching_new}
\end{figure}

Starting with a stable HV state a circularly polarized coherent control pulse, carrying topological charge $m_c{=}{\pm} 1$, is applied to the FM component. This pulse imprints a vortex into the initial FM component [right panel in Fig.~\ref{fig:switching_new}(a)] and simultaneously induces an attractive potential for the other component due to the attractive cross-interaction $-g_{\mathrm{x}}|\Psi_{\mp}|^2$. The density of the initial vortex in the cross-polarized component becomes broader and the balance between the condensate and reservoir, which is key for the vortex stability, is broken. An example of the effective potential landscape during the switching is shown in Fig.~\ref{fig:potential}. As a result, the initial vortex is no longer confined and spirals out of the condensate due to the pulse induced imbalance [left panels in Figs.~\ref{fig:switching_new}(b) and \ref{fig:switching_new}(c)]. The switching does not depend on the total OAM transferred by the resonant pulse as long as the imprinted vortex is stable (after the pulse is gone) and there is no OAM conservation rule actively acting~\cite{PhysRevB.87.081309,PhysRevLett.104.126402}. Important for the switching is the shape and amount of the cross-polarized density, since the destabilization and removal of the vortex in the other component is solely caused by the transient dynamics in the effective potential landscape while the control pulse is present. If the intensity and radius of the control pulse are sufficiently large the initial vortex in the cross-polarized component becomes unstable and leaves the condensate. Effectively, this component transitions into the ground state (FM), while a vortex is imprinted in the other component. This switching process results in the reversal of the circular polarization in the HV core while the OAM is switched from one polarization component to the other one. If the intensity or the duration of the control pulse are not sufficient, it just imprints its vortex state while the other component stays the same, i.e. a FV or SV state with a nonzero topological charge in both components is created. Figure \ref{fig:phase_diagram} shows the time-integrated density induced by the resonant pulse normalized to the time-integrated density induced by the non-resonant pump in order to observe the HV vortex core switching in dependence of the cross-/co-polarized interaction strength ratio. A lower cross-interaction strength can be compensated by a higher pulse power which itself generates a higher cross-polarized density.
\begin{figure}[t]
		\centering
		\includegraphics[width=0.47 \textwidth]{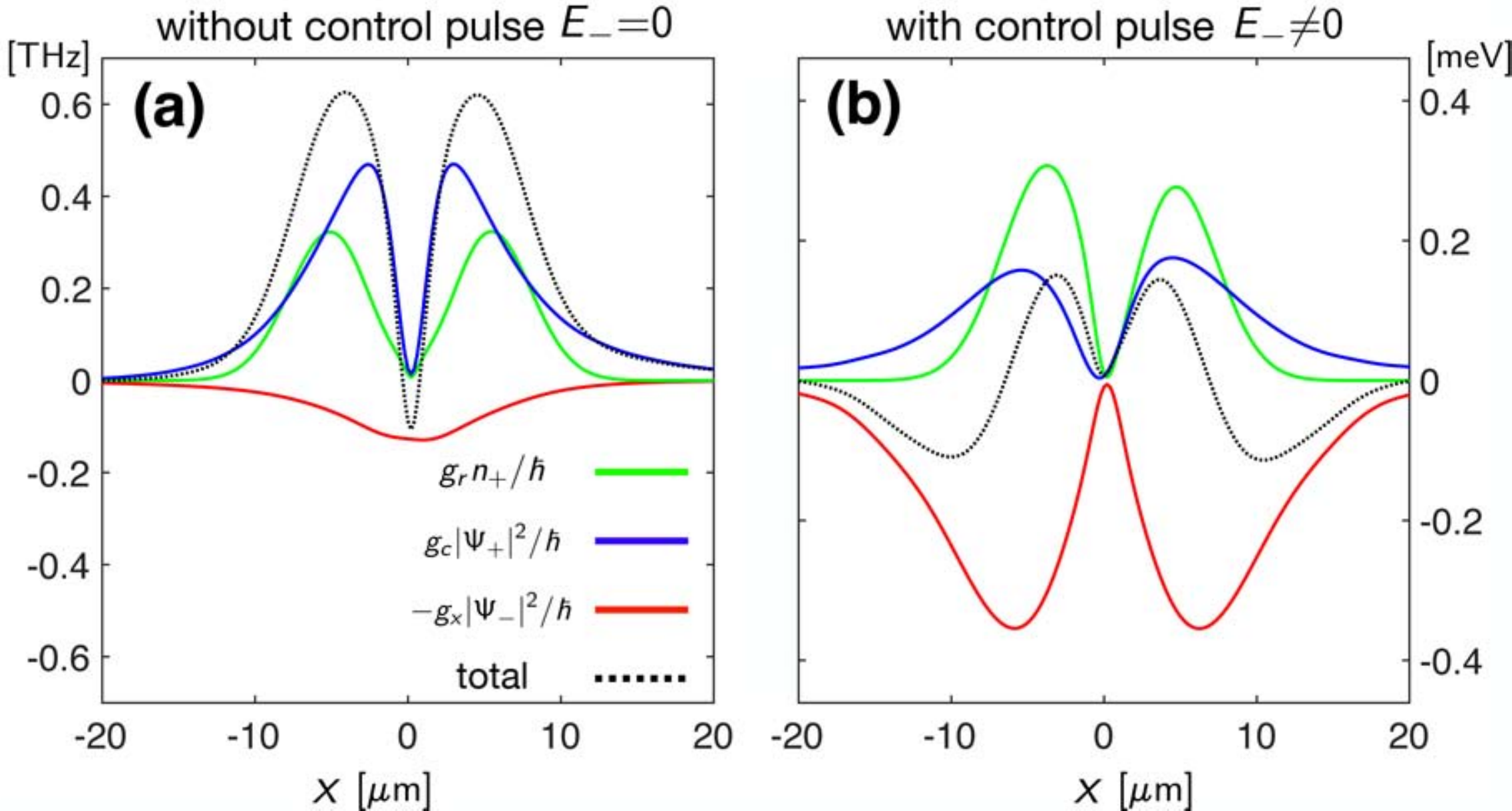}
		\caption{Distributions of the reservoir induced potential (\textit{green}), repulsive self-interaction (\textit{blue}), and attractive cross-interaction (\textit{red}) experienced by the $\Psi_+$ component for the switching example shown in Fig. \ref{fig:HV_switching_example} and \ref{fig:switching_new}. \textit{Black dashed} line shows the total effective potential (sum of the three parts). (a) Without control pulse $E_{-}{=}0$. (b) With control pulse $E_{-}{\neq}0$ and at the time where the maximum intensity is reached [Fig. \ref{fig:switching_new}(b)].}
		\label{fig:potential}
\end{figure}
\begin{figure}[t]
		\centering
		\includegraphics[width=0.38 \textwidth]{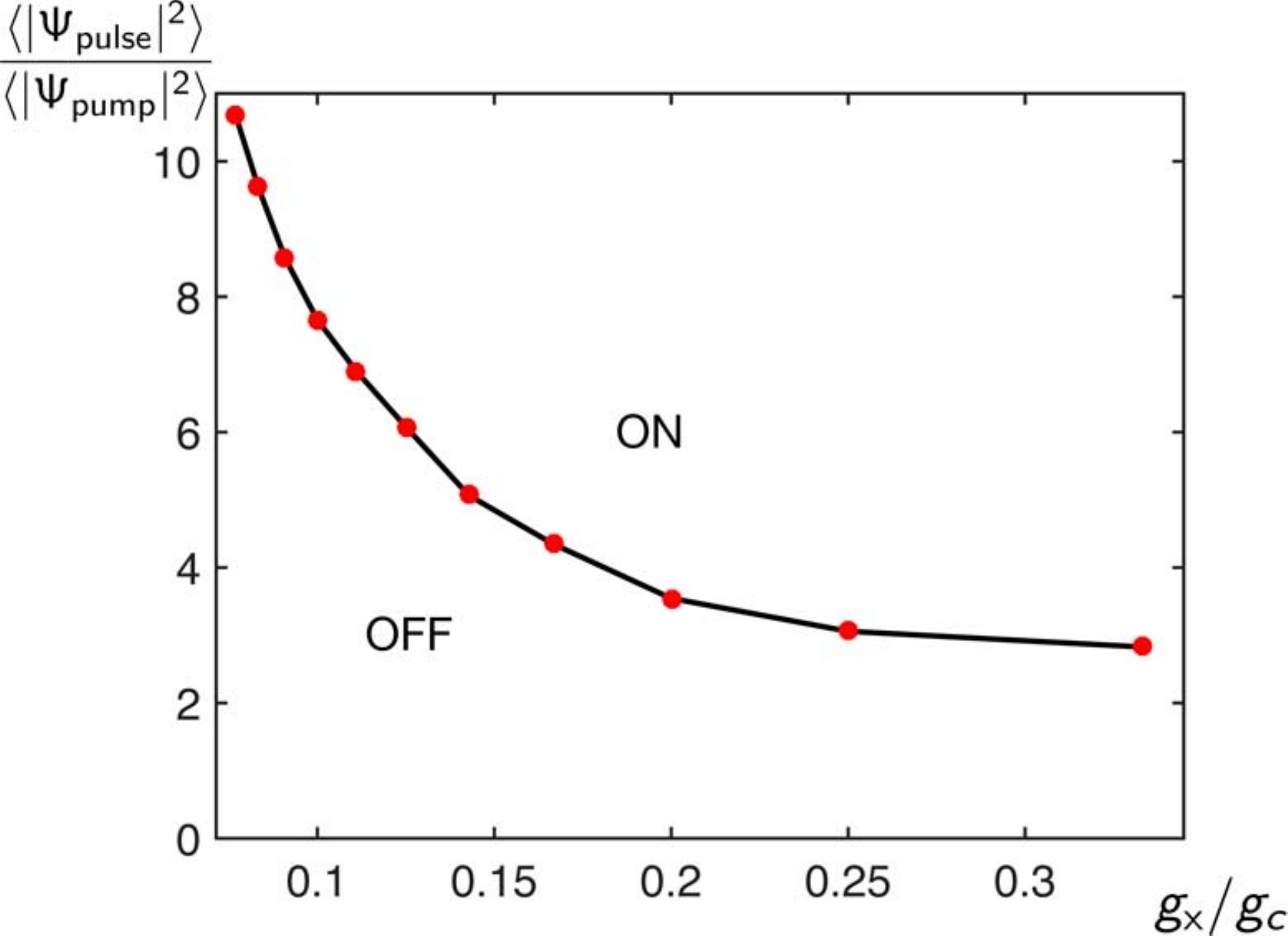}
		\caption{Time-integrated density induced by the resonant pulse $\langle |\Psi_{\mathrm{pulse}}|^2 \rangle$ normalized to the time-integrated density induced by the non-resonant pump $\langle |\Psi_{\mathrm{pump}}|^2 \rangle$ in order to observe the HV vortex core switching in dependence of the cross-/co-polarized interaction strength ratio. This behavior does not depend on the absolute values of the pulse amplitude and duration, but on the time-integrated pulse-induced density. 'ON' marks the parameter region where the HV switching mechanism works. 'OFF' marks the parameter region where the control pulse just imprints a vortex in the corresponding component, without removing the vortex in the other component, and as a result a FV or SV state is created.}
		\label{fig:phase_diagram}
\end{figure}

Figure \ref{fig:HV_switching} shows the possible transitions between the four elementary HVs (\textit{blue solid} arrows) as well as the imprinting of a desired target state (\textit{grey dashed} arrows). Since the switching mechanism is density induced it does not depend on the sign of the topological charge of the control pulse, which means one can switch from a particular left (right) HV to both other right (left) HVs. An example in Fig. \ref{fig:HV_switching_example} shows the density and the phase of the initial HV state $(-1,0)$ [Fig. \ref{fig:HV_switching_example}(a)] and the switched state $(0,-1)$ [Fig. \ref{fig:HV_switching_example}(b)]. In this case, a coherent pulse with $m_c{=}{-}1$ is applied to the $\Psi_{-}$ component, in which the OAM is imprinted (\textit{grey} arrow) and it simultaneously triggers the $\Psi_{+}$ component from $m_{+}{=}{-}1$ to the ground $m_{+}{=}0$ state due to the cross-interaction (\textit{blue} arrow). Figure \ref{fig:potential} shows the effective potential landscape experienced by the initial vortex component $\Psi_+$ during the switching. The pulse induced cross-polarized density suppresses the vortex confinement by the reservoir induced effective potential (cf. Fig. \ref{fig:potential}(b)). For the example shown we use a pulse duration of $\mathrm{w}_t{=}35~\mathrm{ps}$ and the switching process is completed within $150~\mathrm{ps}$. Additionally, the change of density distributions from lemon [Fig. \ref{fig:HV_switching_example}(a)] to star [Fig. \ref{fig:HV_switching_example}(b)] geometry due to finite TE-TM splitting is clearly visible. The corresponding spectra in real space are shown in Figs. \ref{fig:HV_switching_example}(c) and \ref{fig:HV_switching_example}(d), evidencing the vortex core switching process as the simultaneous inversion of two coupled two-level systems. This method presented above provides a platform to perform reversible on-demand all-optical switching between left and right HVs which is largely unaffected by variation of the extrinsic parameters. Another advantage of this vortex core switching process is that it can be detected by simply measuring the polarization resolved intensities in the core region.
\end{section}

\begin{section}{Higher Order States and effect of TE-TM splitting}\label{sec:higher}
\begin{figure}[t]
		\centering
		\includegraphics[width=0.49 \textwidth]{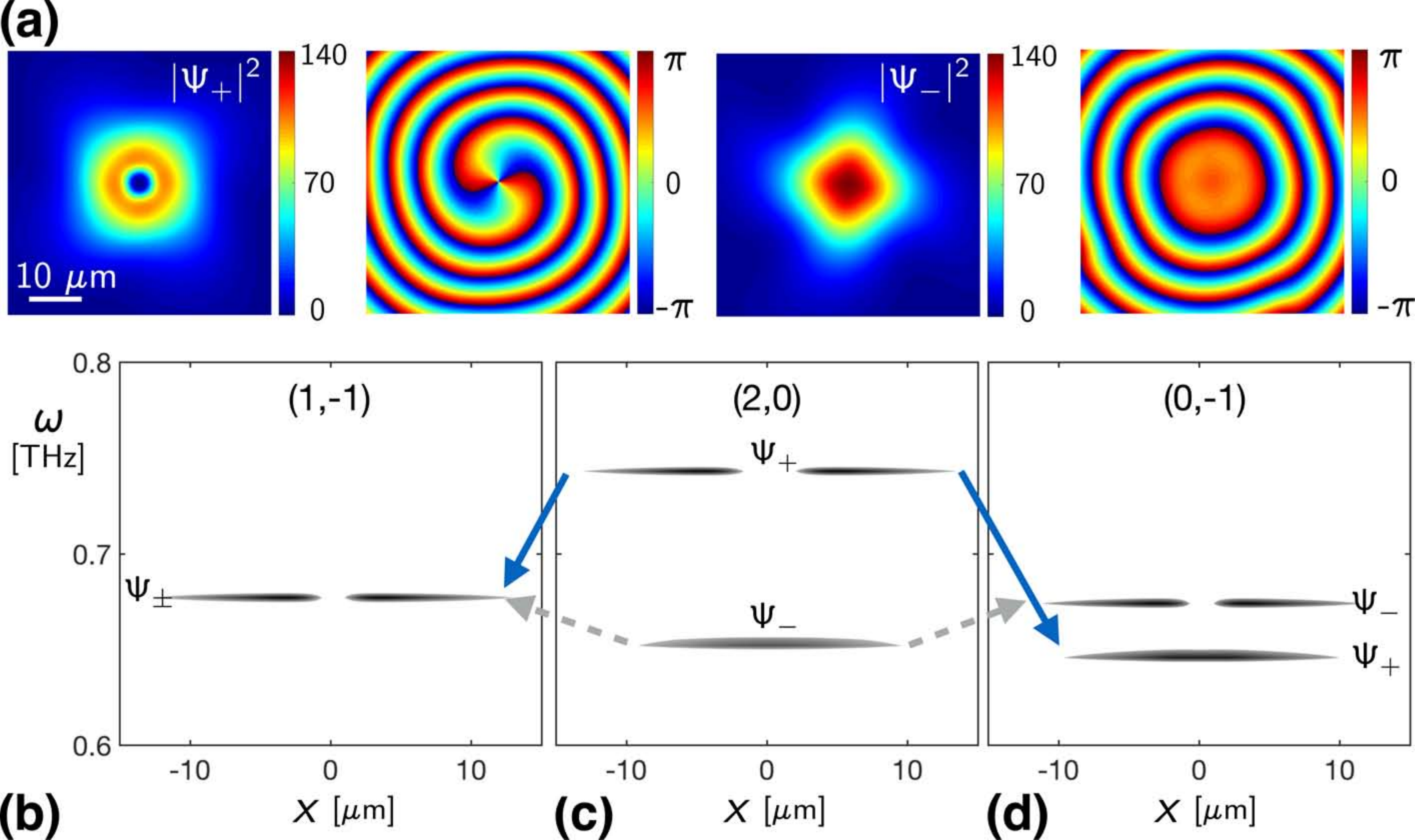}
		\caption{(a) Densities ($\mu m^{{-}2}$) and phase profiles of the higher order HV state $(2,0)$. (c) Spectrum in real space of (a). (b,d) Spectra of possible vortex core switching target states (b) $(1,-1)$ and (d) $(0,-1)$ depending on the intensity of the control beam. Pump radius is $\mathrm{w}_p{=}6~\mu\mathrm{m}$}
		\label{fig:higher_HV_switching}
\end{figure}
Increasing the radius of the pump stabilizes vortex states with higher topological charges $|m_{\pm}|{>}1$ while preserving stability of the ground state and lower order states up to a certain threshold radius \cite{ma2018vortex}. Higher order vortex states with one component in the ground state, i.e. $(m_+,0)$ or $(0,m_-)$, can be exploited for multi-state switching processes. Depending on the intensity and duration of the control pulse, the topological charge of the initial vortex component can be reduced to a lower one or down to the ground state, providing different switching configurations. Starting with a higher order state $(m_+,0)$ with $|m_+|{>}1$, a control pulse with arbitrary topological charge $m_c$ (provided that it is stable), applied to the $\Psi_-$ component, can induce transitions to the state $(m_+',m_c)$ with $|m_+'|{=}|m_+|{-}j$ for $j{=}0$ to $|m_+|$. Figures \ref{fig:higher_HV_switching}(b)-\ref{fig:higher_HV_switching}(d) show the possible vortex core switching from initial state $(2,0)$ to $(1,-1)$ or to $(0,-1)$ as an example using a control pulse with topological charge $m_c{=}{-}1$ applied to the $\Psi_-$ component. Which final state is targeted depends on the intensity and duration of the control pulse. The transient dynamics show that first the $m_+{=}2$ vortex becomes unstable, splits into two $m_+{=}1$ vortices, which then one after the other, spiral out of the condensate. Note that the $|m|{=}2$ vortices are often topologically unstable in experiments, but they are dynamically stable with the density minimum undergoing dynamical splitting and recombination~\cite{dall2014creation,Dominici2018}, which might make the higher-order HV switching easier. In this case, however, the direct measurements of the polarization intensity in the vortex core may not be the method of choice for detection anymore. Instead, OAM-sorting~\cite{ma2019realization} could be used.

\begin{figure}[t]
		\centering
		\includegraphics[width=0.49 \textwidth]{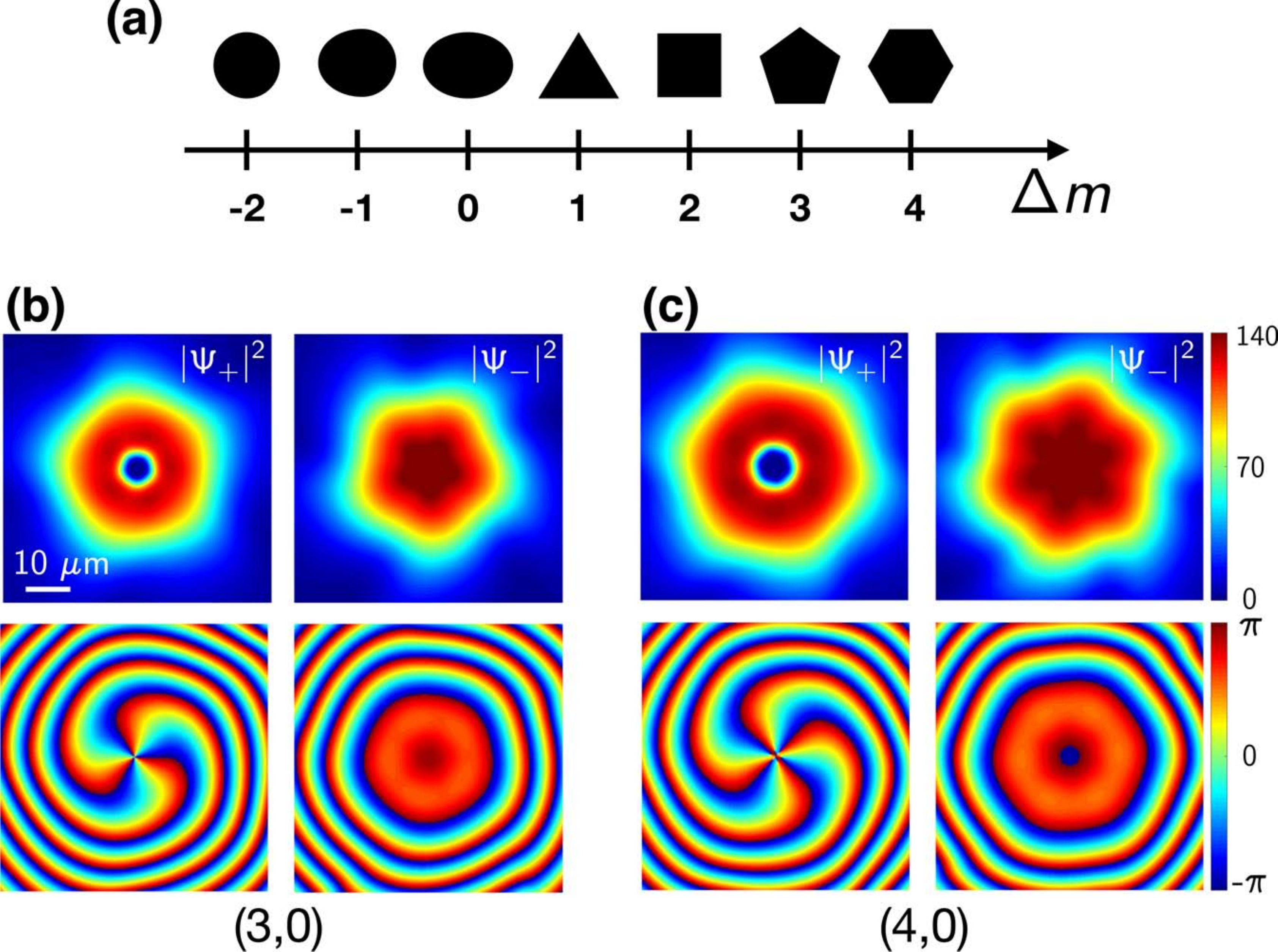}
		\caption{(a) Reshaping of density distributions due to TE-TM splitting depending on the topological charge difference $\Delta m{\equiv} m_+ {-} m_-$. The geometric shapes (from left to right) are circle, lemon, ellipse, star, square, pentagon, and hexagon, respectively. Density and phase profiles of the states (b) $(3,0)$ (pentagon) and (c) $(4,0)$ (hexagon). Pump radii are $\mathrm{w}_p{=}10~\mu\mathrm{m}$ for (b) and $\mathrm{w}_p{=}12~\mu\mathrm{m}$ for (c). The orientations of the patterns rotate over time with periods of about $400~\mathrm{ps}$.}
		\label{fig:fig5}
\end{figure}

For conservative systems in equilibrium without loss and gain, it was shown that including TE-TM splitting leads to cylindrical symmetry breaking, visible as deformations of HV density distributions and phase profiles \cite{solano2010comment,toledo2014warping,gulevich2016topological}. Based on our numerical simulations, we find that this effect is also observed for non-equilibrium systems. Therefore, 
besides having different energies, higher order states can also be sorted into different symmetry groups regarding the approximate geometry of their density distributions. The topological charge difference $\Delta m{\equiv} m_+ {-} m_-$ determines the shape and therefore the corresponding symmetry group which is given by the orthogonal group $O(2)$ for $\Delta m{=}{-}2$ (cylindrical symmetry) and by the dihedral group $D_n$ with $n{=}|\Delta m {+}2|$ otherwise [Fig. \ref{fig:fig5}(a)]. The first $7$ symmetry groups starting from the cylindrical up to hexagonal symmetry are schematically shown in Figs. \ref{fig:fig5}(a). As part of the discussion above, in this paper we explicitly show the following shapes: lemon for $\Delta m{=}{-}1$ [Fig.\ref{fig:HV_switching_example}(a)], star for $\Delta m{=}1$ [Fig. \ref{fig:HV_switching_example}(b)] and square for $\Delta m{=}2$ [Fig.\ref{fig:higher_HV_switching}(a)]. The cylindrically symmetric case $\Delta m {=}{-}2$ and the ellipse for $\Delta m{=}0$ are not shown. Additionally, Fig. \ref{fig:fig5}(b) and \ref{fig:fig5}(c) show the pentagon and hexagon for $\Delta m{=}3,4$. It is worth mentioning that for the higher HV states in Figs. \ref{fig:fig5}(b) and \ref{fig:fig5}(c) the polarization of the vortex core can also be switched by the same method presented above.

\end{section}

\begin{section}{Conclusion}
We have investigated localized HVs in driven-dissipative polariton condensates with finite TE-TM splitting and spin-dependent interaction. The latter is used to realize optical vortex core switching between left and right HVs, which is characterized by circular polarization reversal of the emitted light from the vortex core. In this scenario, only polarization resolved intensity measurement in the core region of the emitted light is required instead of extracting the phase information. We also showed that this switching scheme can be extended to higher order states, leading to multi-state switching configurations.
\end{section}

\begin{acknowledgments}
This work was supported by the Deutsche Forschungsgemeinschaft (DFG) through the collaborative research center TRR142 (grant No. 231447078, project A04) and Heisenberg program (grant No. 270619725) and by the Paderborn Center for Parallel Computing, PC$^2$. X.M. further acknowledges support from the NSFC (No. 11804064).
\end{acknowledgments}

\bibliographystyle{unsrt}
\bibliography{refs}

\begin{thebibliography}{10}
\bibitem{parameters}
The values of the parameters are: $m_{\mathrm{eff}}{=}10^{-4}m_e$ ($m_e$ is the
  free electron mass), $\gamma_r{=}1.5 \gamma_c{=}0.225~\mathrm{ps}^{-1}$,
  $R{=}0.01~\mathrm{ps}^{-1}~\mu\mathrm{m}^2$, $g_r{=}2
  g_c{=}6{\times}10^{-3}~\mathrm{meV}~\mu\mathrm{m}^2$, $g_{\mathrm{x}}{=}0.2
  g_c$, $\Delta_{\mathrm{LT}}{=}0.1~\mathrm{meV}$ (and
  $\Delta_{\mathrm{LT}}{=}0$ for Fig.1),
  $k_{\mathrm{LT}}{=}2~\mu\mathrm{m}^{-1}$. Control beam parameters are:
  $E_0{=}2.5{\times} 10^{-3}~\mu\mathrm{m}^{-3}$,
  $\mathrm{w}_t{=}35~\mathrm{ps}$ for Fig.3,4,5,7b and
  $\mathrm{w}_t{=}15~\mathrm{ps}$ for Fig.7b, $\omega_c{=}0.7~\mathrm{THz}$,
  $\mathrm{w}_c{=}6~\mu\mathrm{m}$ for Fig.3,4,5,6 and
  $\mathrm{w}_c{=}7~\mu\mathrm{m}$ for Fig.7.
\end{thebibliography}

\end{document}